\documentclass[english]{article}
\usepackage[T1]{fontenc}
\usepackage[latin9]{inputenc}
\usepackage[a4paper]{geometry}
\usepackage{color}
\usepackage{float}
\usepackage{calc}
\usepackage{textcomp}
\usepackage{amsmath}
\usepackage{amssymb}
\usepackage{stmaryrd}
\usepackage{graphicx}
\usepackage{setspace}

\makeatletter

\providecommand{\tabularnewline}{\\}

\newcommand{\lyxaddress}[1]{
	\par {\raggedright #1
	\vspace{1.4em}
	\noindent\par}
}

\@ifundefined{date}{}{\date{}}
\makeatother

\usepackage{babel}
\begin{document}

\title{\textbf{\Large{}Exploring quaternion framework for subluminal to
superluminal space transformations in particle dynamics}}

\author{\textbf{\normalsize{}B. C. Chanyal}{\normalsize{}}\thanks{Corresponding Author: email--bcchanyal@gmail.com}{\normalsize{}~}\textsuperscript{1}{\normalsize{},
}\textbf{\normalsize{}L. S. Karki}{\normalsize{}}\textsuperscript{2}{\normalsize{},
}\textbf{\normalsize{}P. K. Joshi}{\normalsize{}}\textsuperscript{3}{\normalsize{},
}\textbf{\normalsize{}B. C. S. Chauhan}{\normalsize{}}\textsuperscript{4}}
\maketitle

\lyxaddress{\begin{center}
\textsuperscript{1,2}\textit{Department of Physics, G. B. Pant University
of Agriculture and Technology, Pantnagar-263 145, (Uttarakhand) India}\\
\textsuperscript{3,4}\textit{Department of Physics, S. S. J. University,
Almora-263 601, (Uttarakhand) India}
\par\end{center}}
\begin{abstract}
The present study explores the behavior of quaternionic four-space
algebra for subluminal and superluminal spaces. We formulate the generalized
Lorentz transformations for quaternionic subluminal, superluminal,
and their combined Minkowski spaces. Furthermore, we have studied
the relativistic phenomenon of quaternionic length contraction, time
dilation, velocity addition, and the Doppler effect for the combination
of subluminal and superluminal space. We claim that the transformation
between two superluminal spaces is ultimately a subluminal space;
the tachyonic behavior reveals itself in the consequences when the
two frames are in different spaces (i.e., generalized subluminal-superluminal
spaces).

\end{abstract}

\section{Introduction:}

~~~~~Since the inception of STR (Special Theory of Relativity)
in 1905 {[}1,2{]}, the study of the dynamic and kinematic behavior
of particles traveling at speeds comparable to the speed of light
has captivated a variety of scientific minds. STR opened new insight
into the understanding of the physical behavior of high-speed particles.
The fundamental postulates of STR established the benchmark of the
theory, which was further perceived in various areas of the scientific
findings. These postulates are further analyzed and extended to incorporate
all types of motions of particles, including the observed and hypothetical
{[}3-7{]}.

The current scenario of the extended version of relativity (STR) classifies
all the observed and hypothetical (which are not experimentally verified
but theoretically possible) particles into two categories {[}5-11{]}
according to their speeds; one is called bradyons (subluminal particles),
which move slower than light, and the other is called tachyons (superluminal
particles), which move faster than light. All experimental findings
to date have resulted from bradyonic particles; therefore, we still
await experimental evidence for tachyonic particles. However, the
very possibility of the mere existence of tachyonic particles has
enthralled many scientific minds. Meanwhile, there exists a well-known
and compatible relativistic theory to describe the dynamic behavior
of particles that are slower than light {[}1--3{]}, but no complete
theory exists for superluminal particles. The Lorentz transformations,
which are the basic building blocks of STR, form themselves a continuous,
non-compact group {[}9,12{]}. The Lorentz group contains not only
the transformation of coordinates due to the relative motions of observers
(called Lorentz boosts) but also the rotation between the other spatial
coordinates. Actually, according to STR, the temporal part of the
motion manifests the same as the spatial part, so the Lorentz boost
is corresponding to the rotation between the time axis and space axis.
Hence, Lorentz transformations correspond to the rotation in four-dimensional
space (1 time and 3 spatial coordinates).

Attempts have been made to incorporate the motion of superluminal
particles within the regime of STR. But this results in causal anomalies
due to the spacelike motion of tachyonic particles {[}6--8{]}.\textcolor{blue}{{}
}In order to establish a fully compatible theory for the description
of the dynamic behavior of superluminal particles, considerable efforts
have been made in the scientific literature. One such significant
attempt has been done {[}3--5{]}, in which the STR is extended to
embrace the dynamics of tachyonic particles by involving the symmetry
arguments within the existing theory. The causal anomalies due to
the space-like motion of tachyonic particles have been eliminated
by beseeching the relativistic \textit{St$\ddot{u}$ckelberg-Feynman}
switching procedure. The formalism involves the symmetric switching
of space and time intervals for tachyonic particles; hence, the space
is extended to six dimensions $\mathbb{R}^{3}\times\mathrm{T}^{3}$.

Recami {[}3,4{]} adopted the first approach, in which the components
of a four-vector field orthogonal to the direction of relative motion
transition from real to imaginary values as they move from subluminal
to superluminal domains. On the other hand, Antippa and Everett {[}6{]}
followed the second approach, utilizing the real superluminal Lorentz
transformation for their analysis. The ultimate speed limit of the
universe, as dictated by the Lorentz transformation, is the speed
of light, setting a boundary for the fastest achievable velocity.
For tachyons, these equations would be modified because they move
faster than light. This leads to some unusual results, such as time
and space becoming complex quantities {[}4--11{]}. The occurrence
of imaginary quantities in the relativistically consistent models
of superluminal particles led some authors {[}9--18{]} to use non-real
numbers (or complex and hyper-complex numbers) in the foundation of
the model.

Teli {[}7{]} tried to unify the quaternionic transformations, which
remain consistent for both subluminal and superluminal Lorentz transformations.
The special theory of relativity has been straightforwardly extended
to encompass superluminal inertial frames, revealing that the presence
of tachyons (particles traveling beyond the speed of light) does not
inherently violate the principles of relativistic theory. When deriving
the superluminal Lorentz transformation for frames with relative velocities
surpassing the speed of light, distinct methodologies have been employed
by various researchers. In the formulation of superluminal Lorentz
transformation {[}8{]}, it has been pointed out that the space for
superluminal particles becomes generalized, called $\mathbb{T}$$^{4}-$space,
in which the roles of space and time are interchanged. Accordingly,
the superluminal transformation {[}10{]} also leads to chronological
mapping $(3,1)\leftrightarrow(1,3).$ Further, Gunawant and Rajpoot
{[}11{]} showed that the components of space and time on passing from
subluminal to superluminal regions can be described as $(-ict,x,y,z)$
$\rightarrow$ ($t$,$-\frac{i}{c}x$,$-\frac{i}{c}y$, $-\frac{i}{c}z$
). Moreover, the role of quaternionic algebra {[}10--27{]} is quite
interesting to construct a generalized relativistic model in modern
physics. Meanwhile, other attempts {[}28--34{]} have been undertaken
for constructing a relativistic and non-relativistic model of several
branches of physics using hyper-complex numbers.

Keeping in view the realm of relativistic physics and the suitability
of using non-commutative algebras, particularly in dealing with the
Lorentz transformations in four dimensions, in this paper we have
generalized the formulation of quaternionic Lorentz transformation
equations in superluminal space. The paper is basically divided into
6 sections, including the introduction. $Section\;2$ contains the
preliminary mathematical formulations of quaternion algebra. In $section\;3$,
we have established the quaternionic form of subluminal and superluminal
Lorentz transformations for different combinations of spaces. The
$section\;4$ contains the quaternionic formulation of the different
relativistic consequences, i.e., length contraction, time dilation,
and velocity addition. In $section\;5$, we have discussed the quaternionic
generalization of the relativistic Doppler effect. $Section\;6$ contain
discussions and conclusions on the planned work, as well as an exploration
of the scope of future work.

\section{Mathematical preliminaries:}

A quaternionic algebra {[}15--18{]} is a type of extended algebra
characterized by four-dimensional spaces that encompass both scalar
and vector components. Mathematically, the 4-dimensional quaternion
algebra can be represented as, 
\begin{align}
\mathbb{Q}\,:\rightarrow & \,\,\,\,\mathbb{Q}(e_{0},e_{j})=e_{0}q_{0}+\sum_{j=1}^{3}e_{j}q_{j}\,,\,\,\,\,\,\,\forall\,q_{0}\in\mathbb{R},\,q_{j}\in\mathbb{R}^{3}\,,\label{eq:1}
\end{align}
where $(e_{0},e_{1}$, $e_{2}$, $e_{3}$) are the quaternionic basis
elements in which $e_{0}$ denotes the scalar unit and $e_{j}$( for
$j=1,2,3$) denotes the vector unit. The multiplication of two quaternions
is determined by the multiplication of their basis elements. Consequently,
the multiplication rules for quaternion units can be expressed as
follows:

\begin{align}
e_{0}^{2}=\, & e_{0}=\,1,\,\,\,\,\,e_{j}^{2}=\,-1.\nonumber \\
e_{0}e_{j}=\, & e_{j}e_{0}=\,e_{j},\,\,\,e_{i}e_{j}=-\delta_{ij}+\epsilon_{ijk}e_{k},\,\,\,\,(\forall\,i,j,k=1,2,3),\label{eq:2}
\end{align}
where $\delta_{ij}$ is the Kronecker delta and $\epsilon_{ijk}$
is the three index Levi-Civita symbol. Suppose, two quaternions $\mathbb{X}$,$\mathbb{\mathbb{Y}}$
can be written by
\begin{align}
\mathbb{X}\mathbb{\,\,} & =\,\,e_{0}x_{0}\,\,+\,\,e_{1}x_{1}\,\,+\,\,e_{2}x_{2}\,\,+\,\,e_{3}x_{3}\,,\nonumber \\
\mathbb{Y}\mathbb{\,\,} & =\,\,e_{0}y_{0}\,\,+\,\,e_{1}y_{1}\,\,+\,\,e_{2}y_{2}\,\,+\,\,e_{3}y_{3}\,,\label{eq:3}
\end{align}
then, the multiplication of these quaternions can be expressed as
\begin{align}
\mathbb{\mathbb{\mathbb{X}}}\circ\mathbb{Y}= & \,\,e_{0}(x_{0}y_{0}-\vec{x}\cdot\vec{y})+e_{j}\left[(x_{0}\vec{y}+y_{0}\vec{x}+(\vec{x}\times\vec{y})_{j}\right],\,\,\,\,(\forall j=1,2,3)\,.\label{eq:4}
\end{align}
Here, quaternionic multiplication is denoted by `$\circ$', whereas
`$\text{\ensuremath{\cdot}}$' and `$\times$' denote the scalar and
vector products, respectively. In equation (\ref{eq:4}), the factor
$\left(\vec{x}\times\vec{y}\right)$ plays a significant role in the
non-commutative nature, and its role becomes very important for the
characteristics of subluminal and superluminal particles. In this
context, the coefficient $e_{0}$ is simply a scalar quantity that
can be defined within the framework of a quaternionic scalar field.
On the other hand, the coefficient $e_{j}$ can be associated with
a pure quaternionic vector field. Consequently, quaternions exhibit
the associative property under multiplication, such that $\mathbb{X}$($\mathbb{\mathbb{Y}}\mathbb{Z})$
= ($\mathbb{X}$$\mathbb{\mathbb{Y}})\mathbb{Z}$. As such, the quaternionic
conjugate $\overline{\mathbb{Q}}$ can be defined as 

\begin{align}
\overline{\mathbb{Q}}= & \,\,\,\overline{e_{0}q_{0}+e_{1}q_{1}+e_{2}q_{2}+e_{3}q_{3}}\,,\nonumber \\
= & \,\,\,e_{0}q_{0}-e_{1}q_{1}-e_{2}q_{2}-e_{3}q_{3}\,,\nonumber \\
= & \,\,\,e_{0}(\mathbb{Q}_{S})-e_{j}(\mathbb{Q}_{V}).\,\,\,\,(\forall j=1,2,3)\label{eq:5}
\end{align}
The subspace of the unit quaternions, satisfying the condition $|\mathbb{Q}|$
= $1$ for
\begin{align}
\mathbb{Q}\,= & \,\,\,S(q)\cos\theta+V(q)\sin\theta\,\,\equiv\,\,\cos\theta+V(q)\sin\theta,\label{eq:6}
\end{align}
where $S(q)=(1,0,0,0)$ is the scalar part of the unit quaternion,
while $V(q)=(0,e_{1}q_{1},e_{2}q_{2},e_{3}q_{3})$ is the vector part
of the unit quaternion. The norm of the quaternions can be defined
as,
\begin{align}
\mathbb{N}= & \,\,\,\sqrt{\mathbb{\overline{Q}}\circ\mathbb{Q}}\,\,=\,\,\sqrt{q_{0}^{2}+q_{1}^{2}+q_{2}^{2}+q_{3}^{2}}.\label{eq:7}
\end{align}

\section{Quaternionic generalization of Lorentz transformation:}

In order to comprehend how particles and space-time structures behave
in both the subluminal (slower than light) and superluminal (faster
than light) regimes, Lorentz transformations are essential to the
theory of special relativity {[}2,35{]}. Thus, in this context, we
will study the quaternion-based formulation of Lorentz transformations
for subluminal and superluminal regimes.

\subsection{{\Large{}Quaternionic subluminal Lorentz transformation:}}

To formulate the quaternionic Lorentz transformation for subluminal
particles, we first introduce a quaternionic form of the Minkowski
four-dimensional space-time interval in $\mathbb{R}^{4}-$space. For
this, a two-dimensional rotation {[}23{]} by an angle $\theta$ can
be defined by the Euler expression:
\begin{align}
e^{i\theta} & =\,\,\cos\theta+i\sin\theta\,\,.\label{eq:8}
\end{align}
Now, we can generalize the equation (\ref{eq:8}) by using the quaternionic
basis elements as {[}5,7{]}
\begin{align}
u_{j}=\exp\left(e_{j}\frac{\phi}{2}\right)=\cos\frac{\phi}{2}+e_{j}\sin\frac{\phi}{2} & ,\,\,\,\,\,\forall\,\,e_{j}^{2}=-1,\label{eq:9}
\end{align}
where $u_{j}$ is a unit quaternion with $(u_{j})=\sqrt{u_{j}\overline{u_{j}}}=1$
($\overline{u_{j}}$ is the quaternionic conjugate of $u_{j}$), which
initiates rotation around the $e_{j}$ axis by an angle $\phi$. From
the above equation (\ref{eq:9}), we may now transform the quaternionic
variable $q$ into $q^{'}$ \textcolor{black}{as, }
\begin{align}
q:\longrightarrow q^{'} & \,=\,\,u_{j}\circ q\circ\overline{u_{j}}^{\ast},\label{eq:10}
\end{align}
where $\overline{u_{j}}$$^{\ast}$ is the quaternion complex conjugate.
Putting the values of $u_{j}$ and $\overline{u_{j}}$$^{\ast}$ into
equation (\ref{eq:10}), we obtain
\begin{align}
q^{'}= & \,\,\left(e_{0}\cos\frac{\phi}{2}+e_{1}\frac{v_{x}}{v}\sin\frac{\phi}{2}+e_{2}\frac{v_{y}}{v}\sin\frac{\phi}{2}+e_{3}\frac{v_{z}}{v}\sin\frac{\phi}{2}\right)\circ\left(e_{0}q_{0}+e_{1}q_{1}+e_{2}q_{2}+e_{3}q_{3}\right)\circ\nonumber \\
\,\,\,\, & \left(e_{0}\cos\frac{\phi^{*}}{2}-e_{1}\frac{v_{x}}{v}\sin\frac{\phi^{*}}{2}-e_{2}\frac{v_{y}}{v}\sin\frac{\phi^{*}}{2}-e_{3}\frac{v_{z}}{v}\sin\frac{\phi^{*}}{2}\right)\,.\label{eq:11}
\end{align}
Here, we have introduced the velocity components ($v\rightarrow(v_{x},v_{y},v_{z})$)
along with the quaternionic frame. Simplifying the above quaternionic
transformation through a complex angle as $\phi=\eta+i\xi,$ so that
$\text{sin}$$(\frac{\phi}{2}-\frac{\phi}{2}^{\ast})=i\text{sinh}\xi$
and $\text{cos}$$(\frac{\phi}{2}-\frac{\phi}{2}^{\ast})=\text{cosh}\xi$,
where the parametric equations are given by $\text{tanh}\xi=\frac{v}{c}=\beta$,
$\text{cosh}\xi=\gamma,$ and $\text{sinh}\xi=\beta\gamma$ (where
$c$ is the speed of light). Thus, we obtain
\begin{align}
q^{'}\left(-ict^{'},\,x^{'},\,y^{'},\,z^{'}\right)= & \,\,e_{0}\gamma i\left[ct-\frac{1}{c}(\vec{v}.\vec{r})\right]\nonumber \\
+ & e_{1}\gamma\left[x-v_{x}t-\frac{i}{c}(\vec{v}\times\vec{r})_{x}\right]\nonumber \\
+ & e_{2}\gamma\left[y-v_{y}t-\frac{i}{c}(\vec{v}\times\vec{r})_{y}\right]\nonumber \\
+ & e_{3}\gamma\left[z-v_{z}t-\frac{i}{c}(\vec{v}\times\vec{r})_{z}\right]\,,\label{eq:12}
\end{align}
where the quaternionic variables $(q_{0}$,$\,q_{1},\,q_{2}$,$\,q_{3}$)
are replaced by $(-ict,\,x,\,y,\,z$) for the Minkowski space-time
structure. Hence, for a unified structure of quaternions, the above
equation (\ref{eq:12}) may be written in a combined form as
\begin{align}
q^{'}\left(-ict^{'},\,\vec{r}\right) & =\,\,e_{0}\gamma i\left[ct-\frac{1}{c}(\vec{v}.\vec{r})\right]+\gamma e_{j}\left[x_{j}-v_{j}t-\frac{i}{c}(\vec{v}\times\vec{r})_{j}\right],\,\,\,\,\,\,(\forall\,j=1,2,3).\label{eq:13}
\end{align}
Equation (\ref{eq:13}) represents the generalized quaternionic Lorentz
transformation in subluminal domains. The above transformation relations
are identical to the previous results {[}5,8{]} if we restrict the
motion of the particle along the $x-$axis direction only. It is noted
that the quaternionic coefficient $e_{0}$ shows the transformation
of the time component, whereas the quaternionic coefficient $e_{j}$
shows the transformation of space components for bradyonic particles.

\subsection{\textcolor{black}{\Large{}Quaternionic superluminal Lorentz transformation:}}

Since the superluminal frames of reference are those frames that travel
at a speed greater than the speed of light with respect to subluminal
frames. Thus, according to the switching principle, the roles of space
and time are interchanged {[}4,10,11{]} in the case of superluminal
transformations. So the generalized space for superluminal particles
is Minkowski \textbf{$\mathbb{T}^{4}-$}space rather than $\mathbb{R}^{4}-$space.
In other words, the superluminal transformations are led by the chronological
space-time mapping {[}8{]}, i.e., $M:(3,1)\longleftrightarrow(1,3)$.
Keeping in view all these arguments together, we generalize the quaternionic
form of Lorentz transformation for superluminal space as,
\begin{align}
t^{q}:\longrightarrow t^{'q} & \,=\,\,u_{j}\circ t^{q}\circ\overline{u_{j}}^{\ast},\label{eq:14}
\end{align}
where $t^{q}=(t_{0},t_{1},\,t_{2},\,t_{3})$ is quaternionic four-times
in Minkowski \textbf{$\mathbb{T}^{4}-$}space. Now, substituting the
quaternionic value of $u_{j}$ an\textcolor{black}{d }$\overline{u_{j}}$$^{\ast}$
\textcolor{black}{fr}om equation (\ref{eq:9}) in equation (\ref{eq:14}),
we get
\begin{align}
t^{'q}= & \left(e_{0}\cos\frac{\phi}{2}+e_{1}\frac{v_{x}}{v}\sin\frac{\phi}{2}+e_{2}\frac{v_{y}}{v}\sin\frac{\phi}{2}+e_{3}\frac{v_{z}}{v}\sin\frac{\phi}{2}\right)\circ\left(e_{0}t_{0}+e_{1}t_{1}+e_{2}t_{2}+e_{3}t_{3}\right)\circ\nonumber \\
\:\;\; & \left(e_{0}\cos\frac{\phi^{*}}{2}-e_{1}\frac{v_{x}}{v}\sin\frac{\phi^{*}}{2}-e_{2}\frac{v_{y}}{v}\sin\frac{\phi^{*}}{2}-e_{3}\frac{v_{z}}{v}\sin\frac{\phi^{*}}{2}\right).\label{eq:15}
\end{align}
By simplifying the equation (\ref{eq:15}) and replacing quaternionic
variables $(t_{0}$, $t_{1},\,t_{2}$,$\,t_{3})$ by ($t,-ix/c,-iy/c,-iz/c$)
then we obtain,
\begin{align}
t^{'q}\left(t^{'},-\frac{i}{c}x^{'},-\frac{i}{c}y^{'},-\frac{i}{c}z^{'}\right)= & \,\,e_{0}\left(\gamma t-\gamma\frac{xv_{x}}{c^{2}}-\gamma\frac{yv_{y}}{c^{2}}-\gamma\frac{zv_{z}}{c^{2}}\right)\nonumber \\
+ & e_{1}\left(i\gamma\frac{v_{x}t}{c}-i\gamma\frac{x}{c}-\gamma\frac{v_{z}y}{c^{2}}+\gamma\frac{zv_{y}}{c^{2}}\right)\nonumber \\
+ & e_{2}\left(-i\gamma\frac{y}{c}-i\gamma\frac{v_{x}z}{c^{2}}+\gamma\frac{v_{z}x}{c^{2}}+i\gamma\frac{v_{y}t}{c}\right)\nonumber \\
+ & e_{3}\left(i\gamma\frac{v_{z}t}{c}-i\gamma\frac{z}{c}+\gamma\frac{v_{x}y}{c^{2}}-\gamma\frac{xv_{y}}{c^{2}}\right)\,\,,\label{eq:16}
\end{align}
which further can be written in terms of vector notations as
\begin{align}
t^{'q}= & \,\,\gamma e_{0}\left(t-\frac{1}{c^{2}}(\vec{v}.\vec{r})\right)+e_{1}\left(-i\frac{\gamma}{c}(x-v_{x}t)-\frac{\gamma}{c^{2}}(\vec{v}\times\vec{r})_{x}\right)\nonumber \\
+ & e_{2}\left(-i\frac{\gamma}{c}(y-v_{y}t)-\frac{\gamma}{c^{2}}(\vec{v}\times\vec{r})_{y}\right)+e_{3}\left(-\frac{i\gamma}{c}(z-v_{z}t)-\frac{\gamma}{c^{2}}(\vec{v}\times\vec{r})_{z}\right).\label{eq:17}
\end{align}
Hence, by equating the scalar and vector parts of the quaternionic
equation (\ref{eq:17}), we get
\begin{align}
t^{'} & \,\,=\,\,\gamma\Biggl(t-\frac{1}{c^{2}}(\vec{v}.\vec{r})\Biggr)\,\,\,;\,\,\,\,\,\,\,\,\,\,\,\,\,\,\,\,\,\,\,\,\,\,\,\,\,\,\,\,\,\,\,\,\,\,\,\,\,\,(\text{Scalar\,part\,of}\,e_{0})\nonumber \\
t_{x}^{'} & \,\,=\,\,\frac{\gamma}{c}\Biggl(x-v_{x}t\Biggr)+\frac{i\gamma}{c^{2}}\Biggl(\vec{v}\times\vec{r}\Biggr)_{x}\,\,\,;\,\,\,\,\,\,\,(\text{Vector\,part\,of}\,e_{1})\nonumber \\
t_{y}^{'} & \,\,=\,\,\frac{\gamma}{c}\Biggl(y-v_{y}t\Biggr)+\frac{i\gamma}{c^{2}}\Biggl(\vec{v}\times\vec{r}\Biggr)_{y}\,\,\,;\,\,\,\,\,\,(\text{Vector\,part\,of}\,e_{2})\nonumber \\
t_{z}^{'} & \,\,=\,\,\frac{\gamma}{c}\Biggl(z-v_{z}t\Biggr)+\frac{i\gamma}{c^{2}}\Biggl(\vec{v}\times\vec{r}\Biggr)_{z}\,\,\,;\,\,\,\,\,\,\,(\text{Vector\,part\,of}\,e_{3}).\label{eq:18}
\end{align}
These relations represent the generalized quaternionic Lorentz transformation
for superluminal particles. Theoretical constructions and hypothetical
particles such as tachyons provide possible foundations for superluminal
motion. It is important to notice that if restricting the relative
motion of a particle along the $x-$axis, i.e., $\overrightarrow{v}\bigparallel x$,
or $\vec{v}\times\vec{r}=0$, the proposed relations reduce to the
result obtained by $\text{Dave\,\,and\,\,Hines\;(1992)}$ {[}8{]}.
Also, if we replace the space-time mapping as $(x,y,z,t)\rightarrow(t^{'},ix^{'},iy^{\prime},iz^{\prime})$,
and $(x,y,z,-t)\rightarrow(t^{\prime},-x^{\prime},-y^{\prime},-z^{\prime})$,
the result coincides with $\text{Recami\:and\;Mignani\;(1974)}$ {[}5{]}. 

As such, we have calculated the quaternionic Lorentz transformation
for a mixed spaces, in which we consider one subluminal space and
other is superluminal space. Here, we can summarize the generalized
quaternionic subluminal and superluminal Lorentz transformation equations
in Table $1$.

\begin{table}[H]
\begin{doublespace}
\centering{}%
\begin{tabular}{ccc}
\hline 
$\mathbb{Q}:\rightarrow\,\,\mathbb{R}^{4}-\mathbb{R}^{'4}$  & $\mathbb{Q}:\rightarrow\,\,\mathbb{T}^{4}-\mathbb{T}^{'4}$  &  $\mathbb{Q}:\rightarrow\,\,\mathbb{R}^{4}-\mathbb{T}^{'4}$ \tabularnewline
\hline 
\hline 
$x^{'}=\gamma\left[(x-v_{x}t)-\frac{i}{c}(\vec{v}\times\vec{r})_{x}\right]$ & $x^{'}=\gamma\left[(x-v_{x}t)-\frac{i}{c}(\vec{v}\times\vec{r})_{x}\right]$ & $x^{'}=-i\frac{\gamma}{c}\left[(x-v_{x}t)-\frac{\gamma}{c^{2}}(\vec{v}\times\vec{r})_{x}\right]$\tabularnewline
$y^{'}=\gamma\left[(y-v_{y}t)-\frac{i}{c}(\vec{v}\times\vec{r})_{y}\right]$ & $y^{'}=\gamma\left[(y-v_{y}t)-\frac{i}{c}(\vec{v}\times\vec{r})_{y}\right]$ & $y^{'}=-i\frac{\gamma}{c}\left[(y-v_{y}t)-\frac{\gamma}{c^{2}}(\vec{v}\times\vec{r})_{y}\right]$\tabularnewline
$z^{'}=\gamma\left[(z-v_{z}t)-\frac{i}{c}(\vec{v}\times\vec{r})_{z}\right]$ & $z^{'}=\gamma\left[(z-v_{z}t)-\frac{i}{c}(\vec{v}\times\vec{r})_{z}\right]$ & $z^{'}=-\frac{i\gamma}{c}\left[(z-v_{z}t)-\frac{\gamma}{c^{2}}(\vec{v}\times\vec{r})_{z}\right]$\tabularnewline
{\small{}$t^{'}=-\gamma\left(t-\frac{1}{c^{2}}(\vec{v}.\vec{r})\right)$} & {\small{}$t^{'}$$=\gamma\left(t-\frac{1}{c^{2}}(\vec{v}.\vec{r})\right)$} & $t^{'}=\frac{i\gamma}{c}\left(t-\frac{1}{c^{2}}(\vec{v}.\vec{r})\right)$\tabularnewline
\hline 
\end{tabular}\caption{Quaternionic transformation equations for $\mathbb{R}^{4}-\mathbb{R}^{'4},\,\mathbb{T}^{4}-\mathbb{T}^{'4}$
and $\mathbb{R}^{4}-\mathbb{T}^{'4}$ spaces}
\end{doublespace}
\end{table}

As mentioned in Table 1, it should be noticed that the generalized
quaternionic Lorentz transformation for $\mathbb{R}^{4}-\mathbb{R}^{'4}$
and $\mathbb{T}^{4}-\mathbb{T}^{'4}$ frames (or within the same frame)
shows the conventional Lorentz transformation equations for bradyonic
particles, while, on the other hand, the generalized quaternionic
Lorentz transformation for $\mathbb{R}^{4}-\mathbb{T}^{'4}$ frame
shows a slightly modified transformation from the conventional one,
which represents the tachyonic behavior of particles.

\section{\textcolor{black}{Consequences of quaternionic Lorentz transformation:}}

In this section we study the very familiar consequences of the special
theory of relativity, like length contraction, time dilation, and
velocity addition in the case of quaternionic Lorentz transformations
for $\mathbb{R}^{4}$ and $\mathbb{T}^{4}-$ spaces.

\subsection{\textcolor{black}{Quaternionic Length contraction:}}

The phenomenon of length contraction in a quaternionic relativistic
model may be derived by assuming two inertial frames of reference:
one $\mathbb{R}^{'4}$ that is the rest frame of the rod, and another
one is $\mathbb{R}^{4}$ frame of reference, which is in relative
motion with respect to $\mathbb{R}^{'4}$ space by a constant velocity
$v$. Now suppose $L_{p}$ is the proper length of the rod in $\mathbb{R}^{'4}$
flat space and $L$ is the length in the $\mathbb{R}^{4}$ flat space,
then the relation between these two lengths in quaternionic form is
defined as
\begin{align}
L\,\,=\,\, & \frac{L_{p}}{\gamma}\,\,,\label{eq:19}
\end{align}
where $L_{p}=e_{0}L_{0}+e_{1}L_{1}+e_{2}L_{2}+e_{3}L_{3}$ is the
quaternionic proper length in $\mathbb{R}^{4}-$ frame. Suppose, for
any two quaternionic frames $\mathbb{R}^{4}$--$\mathbb{R}^{'4}$,
the rest frame $\mathbb{R}^{'4}$ of the rod is moving relative to
frame $\mathbb{R}^{4}$ with a velocity that has components $v_{x}$,
$v_{y}$, $v_{z}$ along the $x$, $y$, and $z-$ axes. Also, the
space-time coordinates of the two ends of the rod \textbf{$AB$} be
taken as ($ict_{1}^{'},x_{1}^{'},y_{1}',z_{1}'$) and ($ict_{2}^{'},x_{2}^{'},y_{2}',z_{2}'$)
respectively, in $\mathbb{R}^{'4}-$frame (see Fig.$1$). The quaternionic
proper length in frame $\mathbb{R}^{'4}$ is then expressed as
\begin{align}
L_{p}^{'} & :\longmapsto\left(e_{0}L_{0}^{'},e_{j}L_{j}^{'}\right)=e_{0}ic\left(t_{2}^{'}-t_{1}^{'}\right)+\sum_{j}e_{j}\left(\vec{r_{2}}^{'}-\vec{r_{1}}^{'}\right)\,\,,\,\,\,\,\,\,\,\,(\forall\,\,j=1,2,3)\label{eq:20}
\end{align}
where $L_{0}^{'}$ is quaternionic scalar length, while the components
of quaternionic vector length $L_{x}^{'}$, $L_{y}^{'}$, and $L_{z}^{'}$
are defined as $\left(\vec{x_{2}}^{'}-\vec{x_{1}}^{'}\right)$, $\left(\vec{y_{2}}^{'}-\vec{y}_{1}^{'}\right)$
and $\left(\vec{z_{2}}^{'}-\vec{z_{1}}^{'}\right)$, respectively.
To find the proper length in $\mathbb{R}^{4}-$frame, we may apply
the quaternionic Lorentz transformation in equation (\ref{eq:20})
with the condition $t_{1}^{'}=t_{2}^{'}$, and obtain
\begin{align}
L_{p}^{'}= & \,\,e_{1}\gamma\left[(\vec{x_{2}}-\vec{x_{1}})-\frac{i}{c}\vec{v}\times(\vec{x_{2}}-\vec{x_{1}})\right]\nonumber \\
+ & e_{2}\gamma\left[(\vec{y_{2}}-\vec{y}_{1})-\frac{i}{c}\vec{v}\times(\vec{y_{2}}-\vec{y}_{1})\right]\nonumber \\
+ & e_{3}\gamma\left[(\vec{z_{2}}-\vec{z_{1}})-\frac{i}{c}\vec{v}\times(\vec{z_{2}}-\vec{z_{1}})\right],\label{eq:21}
\end{align}
which gives,
\begin{align}
L_{p}^{'} & \,\,=\,\gamma\left[(e_{1}L_{x}+e_{2}L_{y}+e_{3}L_{z})-\frac{i}{c}\vec{v}\times(e_{1}L_{x}+e_{2}L_{y}+e_{3}L_{z})\right].\label{eq:22}
\end{align}
Hence, from equation (\ref{eq:27}), we can express the generalized
quaternionic proper length in $\mathbb{R}^{4}-$frame as,
\begin{align}
L_{p}^{'}:\rightarrow\vec{L}_{p}^{'} & \,\,=\gamma\left[\vec{L}-\frac{i}{c}\left(\vec{v}\times\vec{L}\right)\right].\label{eq:23}
\end{align}
Interestingly, the quaternionic proper length suggested by Alam {[}22{]}
is comparable to the one discussed above. Moreover, if the direction
of the length measured is parallel to the direction of velocity, then
$\vec{v}\times\vec{L}=0$ which shows the conventional length contraction.

On the other side, if we consider two quaternionic superluminal frames
$\mathbb{T}^{4}-$ and $\mathbb{T}^{'4}$ instead of real frames for
tachyonic space, where the quaternionic coordinates at the endpoints
of the rod are $t_{A}^{'}:\longrightarrow\left(t_{1}^{'},\,\frac{i}{c}x_{1}^{'},\,\frac{i}{c}y_{1}^{'},\,\frac{i}{c}z_{1}^{'}\right)$
and $t_{B}^{'}:\longrightarrow\left(t_{2}^{'},\,\frac{i}{c}x_{2}^{'},\,\frac{i}{c}y_{2}^{'},\,\frac{i}{c}z_{2}^{'}\right),$then
the quaternionic proper length in $\mathbb{T}^{'4}-$frame becomes
\begin{align}
cT_{p}^{'} & =c\left(t_{B}^{'}-t_{A}^{'}\right)=e_{1}i(x_{2}^{'}-x_{1}^{'})+e_{2}i(y_{2}^{'}-y_{1}^{'})+e_{3}i(z_{2}^{'}-z_{1}^{'})\,,\label{eq:24}
\end{align}
where $t_{1}^{'}=t_{2}^{'}$, and $T_{p}^{'}$ is the quaternionic
proper time in tachyonic space. 

\noindent\fbox{\begin{minipage}[t]{1\columnwidth - 2\fboxsep - 2\fboxrule}%
\begin{figure}[H]
\centering{}\includegraphics[scale=0.45]{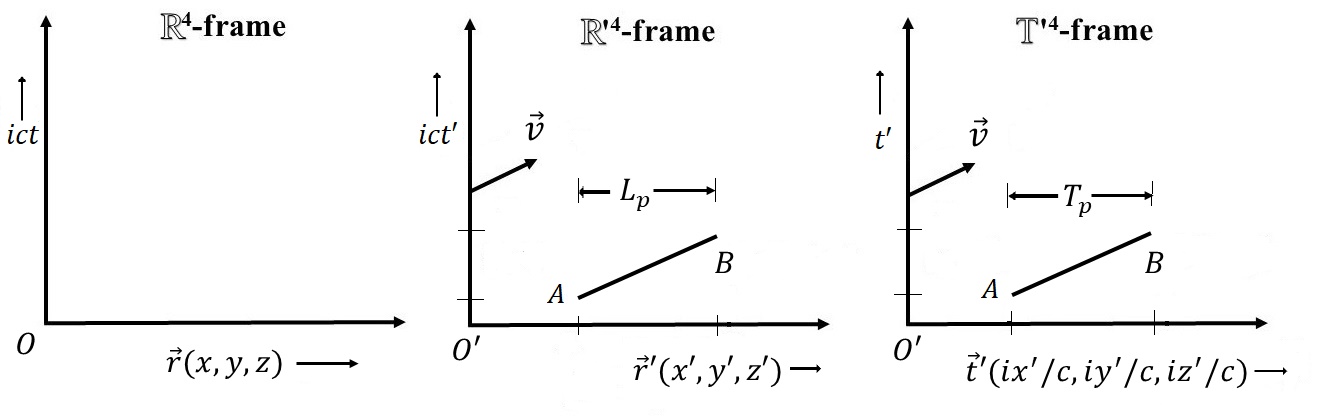}\caption{Length contraction in quaternionic four-space}
\end{figure}
\end{minipage}}

Now, by using the quaternionic superluminal Lorentz transformation,
we get,
\begin{align}
T_{p}^{'}= & \frac{\gamma}{c}e_{1}\left[(\vec{x_{2}}-\vec{x_{1}})-\frac{i}{c}\vec{v}\times(\vec{x_{2}}-\vec{x_{1}})\right]\nonumber \\
+ & \frac{\gamma}{c}e_{2}\left[(\vec{y_{2}}-\vec{y_{1}})-\frac{i}{c}\vec{v}\times(\vec{y_{2}}-\vec{y_{1}})\right]\nonumber \\
+ & \frac{\gamma}{c}e_{3}\left[(\vec{z_{2}}-\vec{z_{1}})-\frac{i}{c}\vec{v}\times(\vec{z_{2}}-\vec{z_{1}})\right],\label{eq:25}
\end{align}
which can be further written as
\begin{align}
T_{p}^{'}= & \frac{\gamma}{c}\left[(e_{1}L_{x}+e_{2}L_{y}+e_{3}L_{z})-\frac{i}{c}\vec{v}\times(e_{1}L_{x}+e_{2}L_{y}+e_{3}L_{z})\right]\,.\label{eq:26}
\end{align}
Hence, from equation (\ref{eq:26}) the proper length in quaternionic
$\mathbb{T}^{4}-$space becomes
\begin{align}
\vec{L}_{p}^{'} & \,\,\equiv\,\,cT_{p}^{'}\,=\,\,\gamma\left(\vec{L}-\frac{i}{c}(\vec{v}\times\vec{L})\right).\label{eq:27}
\end{align}
This is the quaternionic form of length contraction for superluminal
$\mathbb{T}^{4}-$space. Theoretically, the quaternionic length contraction
exhibits a complex nature for tachyons. If we choose $\vec{v}\parallel\vec{L}$
or $\vec{v}\times\vec{L}=0$, it reduces to $\vec{L}_{p}^{'}=\gamma\vec{L}.$
Thus, the quaternionic length contraction evaluated for the case of
superluminal spaces shows the same form as derived in $\mathbb{R}^{4}-$spaces.

However, if we consider two quaternionic frames in different spaces,
i.e., $\mathbb{R}^{4}$ and $\mathbb{T}^{'4}$--spaces, in which
$\mathbb{T}^{'4}$ is in the rest frame of the rod, moving with velocity
$v$ w.r.t. $\mathbb{R}^{4}$ frame (see Fig.$1$). Then, the quaternionic
rest time in $\mathbb{T}^{'4}$ frame can be expressed as
\begin{align}
T_{p}^{'} & \,\,=\left(t_{A}^{'}-t_{B}^{'}\right)=e_{1}\frac{i}{c}(x_{2}^{'}-\vec{x_{1}}^{'})+e_{2}\frac{i}{c}(y_{2}^{'}-y_{1}^{'})+e_{3}\frac{i}{c}(z_{2}^{'}-z_{1}^{'})\,,\label{eq:28}
\end{align}
where $t_{A}^{'}:\longrightarrow\left(t_{1}^{'},\,\frac{i}{c}x_{1}^{'},\,\frac{i}{c}y_{1}^{'},\,\frac{i}{c}z_{1}^{'}\right)$
and $t_{B}^{'}:\longrightarrow\left(t_{2}^{'},\,\frac{i}{c}x_{2}^{'},\,\frac{i}{c}y_{2}^{'},\,\frac{i}{c}z_{2}^{'}\right)$
are two ends of the rod with $t_{1}^{'}=t_{2}^{'}$. Now, by using
quaternionic superluminal Lorentz transformations, we obtain

\begin{align}
T_{p}^{'}= & \,\,e_{1}i\frac{\gamma}{c}\left[(\vec{x_{2}}-\vec{x_{1}})-\frac{i}{c}\vec{v}\times(\vec{x_{2}}-\vec{x_{1}})\right]\nonumber \\
+ & e_{2}i\frac{\gamma}{c}\left[(\vec{y_{2}}-\vec{y_{1}})-\frac{i}{c}\vec{v}\times(\vec{y_{2}}-\vec{y_{1}})\right]\nonumber \\
+ & e_{3}i\frac{\gamma}{c}\left[(\vec{z_{2}}-\vec{z_{1}})-\frac{i}{c}\vec{v}\times(\vec{z_{2}}-\vec{z_{1}})\right]\,\,.\label{eq:29}
\end{align}
Simplifying equation (\ref{eq:29}) as
\begin{align}
T_{p}^{'}= & \,\,\frac{i\gamma}{c}\left[(e_{1}L_{x}+e_{2}L_{y}+e_{3}L_{z})-\frac{i}{c}\vec{v}\times(e_{1}L_{x}+e_{2}L_{y}+e_{3}L_{z})\right]\,\,,\label{eq:30}
\end{align}
which yields the quaternionic proper length as,
\begin{align}
\vec{L}_{p}^{'} & \,\,=i\gamma\left(\vec{L}-i\frac{1}{c}(\vec{v}\times\vec{L})\right)\,\,.\label{eq:31}
\end{align}
In case, $\vec{v}\times\vec{L}=0$, then the quaternionic length contraction
measured from subluminal to superluminal space (i.e., $\mathbb{R}^{4}$--$\mathbb{T}^{'4}$
transformation) of tachyons becomes $L=-i\left(\sqrt{\frac{v^{2}}{c^{2}}-1}\right)\vec{L}_{p}^{'}$,
which shows imaginary for tachyonic particles.

\subsection{\textcolor{black}{Quaternionic Time dilation:}}

Time dilation is a relativistic phenomenon in which a moving object
experiences time slower than a stationary observer. Thus, in this
context, we formulate the quaternionic form of time dilation for considering
two frames, i.e., $\mathbb{R}^{4}$ and $\mathbb{T}^{4}-$ frames.
Let us start with a simple case for two quaternionic frames as $\mathbb{R}^{4}$
and $\mathbb{R}^{'4}-$frames, where $\mathbb{R}^{'4}-$frame is moving
with velocity $v$ relative to frame $\mathbb{R}^{4}$ (see Fig.2).
Suppose a clock is at rest in the moving frame $\mathbb{R}^{'4}$,
positioned at $\vec{r}$ from the origin $\text{O}$$^{'}$. The quaternionic
proper time in frame $\mathbb{R}^{'4}$ is given by the time interval
between the events, i.e.
\begin{align}
T_{p} & \,\,=e_{0}ic(t_{2}^{'}-t_{1}^{'})+e_{1}(x_{2}^{'}-x_{1}^{'})+e_{2}(y_{2}^{'}-y_{1}^{'})+e_{3}(z_{2}^{'}-z_{1}')\,\,.\label{eq:32}
\end{align}

\noindent\fbox{\begin{minipage}[t]{1\columnwidth - 2\fboxsep - 2\fboxrule}%
\begin{figure}[H]
\centering{}\includegraphics[scale=0.4]{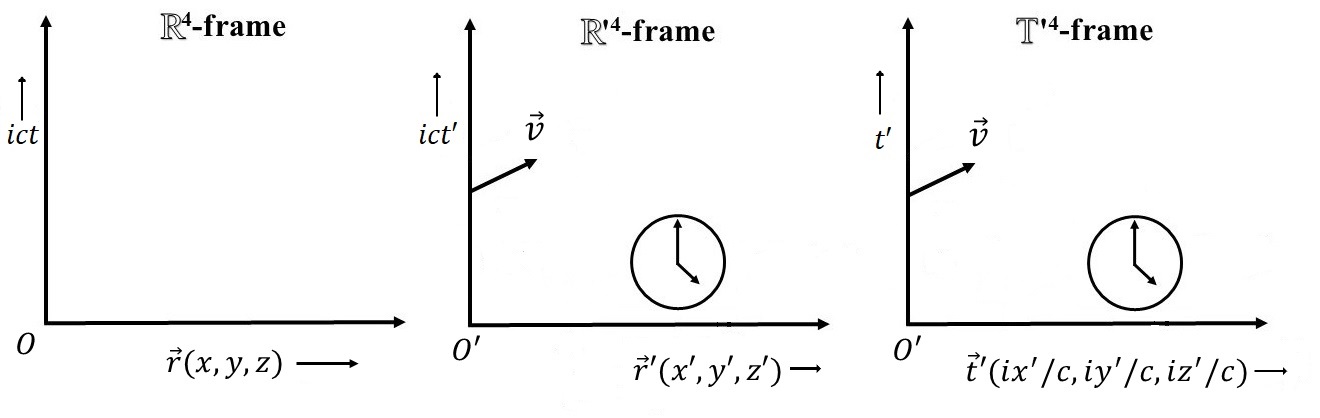}\caption{Time dilation in quaternionic four-space}
\end{figure}
\end{minipage}}

Since the clock is at rest in $\mathbb{R}^{'4}-$frame, so that $x_{j+1}^{'}=x_{j}^{'}$,
$y_{j+1}^{'}=y_{j}^{'}$, and $z_{j+1}^{'}=z_{j}^{'}$, $\forall$$j=(1,2,3)$.
Thus,
\begin{align}
T_{p} & \,\,=\,\,ic(t_{2}^{'}-t_{1}^{'}).\,\,\,\,\,(\text{Scalar\,coefficient\,of\,\ensuremath{e_{0}}}).\label{eq:33}
\end{align}
Using quaternionic inverse Lorentz transformations, we can express
the time dilation {[}22{]} for $\mathbb{R}^{4}-$ frame as
\begin{align}
T\,= & \,\,\left(t_{2}-t_{1}\right)\,=\,-\gamma ic\left(t_{2}^{'}+\frac{1}{c^{2}}(\vec{v}.\vec{r}^{'})\right)+\gamma ic\left(t_{1}^{'}+\frac{1}{c^{2}}(\vec{v}.\vec{r}^{'})\right)\,,\nonumber \\
= & \,-\gamma ic\left(t_{2}^{'}-t_{1}^{'}\right)\,=\,-\gamma T_{p}\,\,.\,\,\,\,\,\,\,\,\,\,(\text{for \ensuremath{\mathbb{R}^{4}-\mathbb{R}^{'4}}}\text{space})\label{eq:34}
\end{align}
On the other hand, if we consider two quaternionic frames, i.e., $\mathbb{T}^{4}$
and $\mathbb{T}^{'4}$, in superluminal spaces, in which a clock at
rest in the moving frame $\mathbb{T}^{'4}$. The quaternionic proper
time in frame $\mathbb{T}^{'4}$ is given by the time interval between
the events, i.e.,
\begin{align}
T_{p} & \,\,=\,\,e_{0}\left(t_{2}^{'}-t_{1}^{'}\right)\,\,.\label{eq:35}
\end{align}
By applying the inverse quaternion superluminal Lorentz transformations,
then the time dilation becomes
\begin{align}
T= & \,\,\left(t_{2}-t_{1}\right)\,\,=\,\,-\gamma\left(t_{2}^{'}+\frac{1}{c^{2}}(\vec{v}.\vec{r}^{'})\right)+\gamma\left(t_{1}^{'}+\frac{1}{c^{2}}(\vec{v}.\vec{r}^{'})\right)\nonumber \\
= & \,\,-\gamma(t_{2}^{'}-t_{1}^{'})\,=\,-\gamma T_{p}.\,\,\,\,\,\,\,\,(\text{for \ensuremath{\mathbb{T}{}^{4}}\ensuremath{\ensuremath{-\mathbb{T}^{'}}\ensuremath{{}^{4}}}}\text{space})\,.\label{eq:36}
\end{align}
Interestingly, the time dilation within the quaternionic two superluminal
spaces appears to be the same as in two subluminal spaces. However,
if we consider two quaternionic distinct frames, $\mathbb{R}^{4}$
and $\mathbb{T}^{'4}$, in which $\mathbb{T}^{'4}-$frame is moving
with a relativistic speed $v$. Suppose a stationary clock $T_{p}$
in $\mathbb{T}^{'4}-$frame, and two events $\left(t_{1}^{'},\frac{ix_{1}^{'}}{c},\frac{iy_{1}^{'}}{c},\frac{iz_{1}^{'}}{c}\right)$
and $\left(t_{2}^{'},\frac{ix_{2}^{'}}{c},\frac{iy_{2}^{'}}{c},\frac{iz_{2}^{'}}{c}\right)$
are measured by clock. Then, the proper time in quaternionic form
becomes
\begin{align}
T_{p} & \,\,=\,\,e_{0}(t_{2}^{'}-t_{1}^{'})+\frac{i}{c}e_{j}(r_{j+1}^{'}-r_{j}^{'}),\,\,\,(\forall\,j=1,2,3).\label{eq:37}
\end{align}
Now if the events occur at the same stationary position in the $\mathbb{T}^{'4}-$frame,
i.e., $r_{j+1}^{'}=r_{j}^{'}$, then
\begin{align}
T_{p} & \,\,=\,\,e_{0}\left(t_{2}^{'}-t_{1}^{'}\right)\,.\label{eq:38}
\end{align}
Correspondingly, if we measure time events in $\mathbb{R}^{4}-$space,
we get
\begin{align}
T\,= & \,\,\left(t_{2}-t_{1}\right)\,\,=\,\,i\gamma T_{p}\,.\,\,\,\,\,\,\,\,\,\,(\text{for \ensuremath{\mathbb{R}{}^{4}-\mathbb{T}^{'}{}^{4}} }\text{space})\label{eq:39}
\end{align}
It should be noticed that the time dilation in $\mathbb{T}^{4}-$space
is imaginary in comparison to $\mathbb{R}^{4}-$space, which shows
the tachyonic behavior of particles.

\subsection{\textcolor{black}{Quaternionic velocity addition:}}

Since the classical addition is no longer acceptable at speeds approaching
the speed of light due to relativistic effects. For instance, in high-energy
physics, the particles in the Large Hadron Collider (LHC) travel at
relativistic speeds. The proper use of relativistic velocity addition
is required to compute the combined velocities of particles in collisions
and to predict the results of particle interactions. Thus, we represent the
quaternionic version of the relativistic velocity addition for subluminal
and superluminal paces.
\begin{verse}
(i) Within the real Minkowski frame, the relativistic velocity addition
by using quaternionic Lorentz transformation in subluminal frames
can be expressed as
\begin{align}
\vec{U_{\mathbb{R}}}\,:=\frac{\,\,\vec{U}^{'}+\vec{v}+\frac{i}{c}(\vec{v}\times\vec{U}^{'})}{\left(1+\frac{\vec{U}^{'}\cdot\vec{v}}{c^{2}}\right)}\,\,\,,\,\,\,\,\,\,\,\,\,\,(\text{velocity addition in \ensuremath{\mathbb{R}{}^{4}-\mathbb{R}^{'}{}^{4}} }\text{space} & )\label{eq:40}
\end{align}
where $\vec{U}^{'}$ is the velocity of particle in moving frame $\mathbb{R}^{'}{}^{4}$,
$\vec{v}$ is the velocity of moving frame, and $\vec{U_{\mathbb{R}}}$
is the velocity of particle in frame $\mathbb{R}^{4}$.

(ii) Within the pure superluminal frame, we can write the quaternionic
velocity addition as
\begin{align}
\vec{U}_{\mathbb{T}}\,:= & \,\,\frac{\vec{U}^{'}+\vec{v}+\frac{i}{c}(\vec{v}\times\vec{U}^{'})}{\left(1+\frac{\vec{U}^{'}\cdot\vec{v}}{c^{2}}\right)}\,\,\,,\,\,\,\,\,\,\,\,\,\,(\text{velocity addition in \ensuremath{\mathbb{T}{}^{4}-\mathbb{T}^{'}{}^{4}} }\text{space})\,.\label{eq:41}
\end{align}

(iii) For two distinct frames, the quaternionic velocity addition
can be represented as

\begin{align}
\vec{U_{\mathbb{R}}}:=\, & \,-\frac{[\vec{U}^{'}+\vec{v}+\frac{i}{c^{2}}(\vec{v}\times\vec{U}^{'})]}{\left(1+\frac{\vec{U}^{'}\cdot\vec{v}}{c^{2}}\right)}\,\,\,,\,\,\,\,\,\,\,\,\,\,(\text{velocity addition in \ensuremath{\mathbb{R}{}^{4}-\mathbb{T}^{'}{}^{4}} }\text{space}).\label{eq:42}
\end{align}
Here, the negative sign indicates the opposite direction of velocity
appearing in the rest subluminal frame. 
\end{verse}
Hence, the quaternionic relativistic velocity addition is essential
for understanding the behavior of subluminal and superluminal particles
in accelerators. This formalism ensures the correct combination of
two relativistic velocities.

\section{\textcolor{black}{\large{}Quaternionic relativistic Doppler effect:}}

Let us consider two subluminal frames of reference, i.e., $\mathbb{R}^{4}$
and $\mathbb{R}^{'}{}^{4}$, as discussed in earlier sections. If
two light pulses are transmitted at $t=0$ and $t=T,$then $\Delta$t$^{'}$
be the interval between the reception of these pulses in the $\mathbb{R}^{'}{}^{4}-$frame
represents the proper time interval. Since the receiver remains stationary
in $\mathbb{R}^{'}{}^{4}$, the distance $\Delta r^{'}$ covered by
the receiver in $\mathbb{R}^{'}{}^{4}-$frame during the reception
of the two pulses is zero. Utilizing the inverse quaternionic Lorentz
equation, considering $\Delta$$r^{'}=0$ and $\Delta$$t^{'}=$$T^{'}$,
we may express
\begin{align}
\Delta r\,\,=\,\,\gamma\Delta t^{'}v\,\,\,\equiv\,\,\, & \gamma T^{'}v\,\,.\label{eq:43}
\end{align}
In comparison to the first pulse in frame $\mathbb{R}^{4}$, the second
pulse must travel an additional distance $\Delta r$ in order to reach
the frame $\mathbb{R}^{'}{}^{4}$. So that,
\begin{align}
\Delta t & \,\,\,=\,\gamma\left(\Delta t^{'}+\frac{v\Delta r^{'}}{c^{2}}\right)\,\,.\label{eq:44}
\end{align}
We have $\Delta r^{'}$ = 0 and $\Delta$$t^{'}$= $T^{'}$, thus
$\Delta t\,\,=\,\gamma T^{'}.$ If the actual time period for the
first pulse is $T$, and for the second pulse is ($\frac{\Delta r}{c}$)
to cover additional distance $\Delta r$ in frame $\mathbb{R}^{4}$,
then 
\begin{align}
\Delta t & \,\,\equiv\,\,\left(T+\frac{\Delta r}{c}\right)\,\,.\label{eq:45}
\end{align}
On comparing equations (\ref{eq:44}) and (\ref{eq:45}), we get
\begin{align}
T & \,\,=\,\,\gamma T^{'}\left(1-\frac{v}{c}\right)\,\,.\label{eq:46}
\end{align}
This is the transformation relation for time period of two pulses.
If $\nu$ and $\nu^{'}$ are actual and observed frequencies of light
pulse, respectively, $\nu=\frac{1}{T}$ and $\nu^{'}=\frac{1}{T^{'}}$.
Then, we obtain
\begin{align}
\nu & \,\,=\,\,\nu^{'}\sqrt{\frac{(1+\frac{v}{c})}{(1-\frac{v}{c})}}\,\,.\label{eq:47}
\end{align}
Equation (\ref{eq:47}) indicates the relativistic Doppler effect
in terms of quaternionic form for subluminal spaces. Accordingly,
we obtain the similar frequencies transformation formulation for two
homogeneous superluminal frames (e.g., $\mathbb{T}^{4}$ and $\mathbb{T}^{'4}-$frames).
However, if we consider two distinct frames of references; one is
subluminal $\mathbb{R}^{4}$, and the other one is $\mathbb{T}^{'4}$,
where $\mathbb{T}^{'4}$ is moving with velocity $v$ with respect
to $\mathbb{R}^{4}$, then the frequencies transformation relation
is not identical to that of the homogeneous frames. Hence, we have
calculated the following transformation relation for two frequencies
observed in $\mathbb{R}^{4}$ and $\mathbb{T}^{'4}-$ frames as
\begin{align}
\nu^{'} & \,\,=\,\,i\gamma\nu\left(1+\frac{v}{c^{2}}\right)\,\,.\label{eq:48}
\end{align}
Therefore, it is concluded that the relativistic Doppler effect shows
similar behavior when observed from two tachyonic frames ($\mathbb{T}^{4}$
and $\mathbb{T}^{'4}$) or in two bradyonic spaces ($\mathbb{R}^{4}$
and $\mathbb{R}^{'}{}^{4}$), and it becomes imaginary if we observe
it in mixed spaces such as $\mathbb{R}^{4}$ and $\mathbb{T}^{'4}-$
spaces. On the other hand, we may argue that the transformation between
two superluminal spaces is itself a subluminal space; nonetheless,
the tachyonic behavior appears in the consequences when the two frames
are in distinct spaces.

\section{Conclusions:}

The quaternionic subluminal and superluminal transformations have
been studied extensively by generalizing the definition of two-dimensional
rotation into $4$-dimensional quaternionic rotation. The quaternionic
$\mathbb{R}^{4}$ and $\mathbb{T}^{4}$ spaces have been constructed,
and from therein the Lorentz transformations in subluminal (bradyonic)
and superluminal (tachyonic) spaces have been derived. Equation (\ref{eq:13}),
shows the generalized form of quaternionic subluminal transformations,
which reduces to the result derived by $Lin$ $(2022)$ \cite{key-27}
if the motion is taken along the x-axis only. By applying chronological
mapping as $(3,1)\longleftrightarrow(1,3)$, we have derived the quaternionic
superluminal transformation for a general motion in equation (\ref{eq:17}).
The generalized quaternionic Lorentz transformation between subluminal
and superluminal spaces is established, which has been summarized
in Table 1. These relations are similar to the result derived by $Dave\,\,and\,\,Hines\;(1992)$
{[}8{]} for restricted motion of a particle. Hence, by applying non-commutative
quaternionic algebra, the Lorentz transformations within the bradyonic
space (both observers in subluminal) and tachyonic space (both observers
in superluminal) and in combined space (if one observer is in tachyonic
and another is in bradyonic) have been studied extensively.

After establishing all these quaternionic Lorentz transformations
for different combinations of spaces, we have discussed the consequences
in quaternionic four-dimensional subluminal as well as superluminal
spaces for a general motion. The quaternionic length contraction in
subluminal spaces has been established in equation (\ref{eq:23}),
which coincides with the result derived by $Alam\;(2011$) {[}22{]}.
Similarly, for quaternionic superluminal spaces and subluminal-superluminal
spaces, the length contraction is derived in equations (\ref{eq:27})
and (\ref{eq:31}), respectively. 

The quaternionic time dilation for subluminal spaces has also been
established in equation (\ref{eq:45}), which coincides to the result
given by \cite{key-22}. For the case of superluminal spaces and combined
space (subluminal-superluminal) the results are shown by equations
(\ref{eq:36}) and (\ref{eq:39}), respectively. It has been concluded
that the form of proposed equations in subluminal and superluminal
spaces remains the same, however, an extra imaginary factor occurs
for the case of combined space (subluminal-superluminal), which show
the tachyonic behavior of the particles. Moreover, we have discussed
the quaternionic form of relativistic velocity addition for different
combinations of spaces in equations (\ref{eq:40}) (for subluminal
spaces), (\ref{eq:41}) (for superluminal spaces) and (\ref{eq:42})
(for combined space of sub and superluminal spaces), respectively.
Furthermore, in section 5, we have discussed the quaternionic generalization
of the relativistic Doppler effect for different combinations of spaces.

Hence, it is claimed that the relativistic Doppler effect behaves
similarly when observed from two bradyonic spaces ($\mathbb{R}^{4}$
and $\mathbb{R}^{'}{}^{4}$) or two tachyonic frames ($\mathbb{T}^{4}$
and $\mathbb{T}^{'}{}^{4}$), and that it becomes imaginary when viewed
in mixed spaces ($\mathbb{R}^{4}$ and $\mathbb{T}^{'}{}^{4}-$spaces).
Although we might say that the transition between two superluminal
spaces is ultimately a subluminal space, the tachyonic behavior reveals
itself in the consequences when the two frames are in different spaces
(subluminal-superluminal spaces). Of course, there is no direct experimental
evidence for superluminal travel or combined subluminal-superluminal
spaces. It is important to conclude that the behavior of tachyons
becomes imaginary for many reasons. The mass of tachyons is imaginary,
i.e., $\mathcal{M}=i\mu,$ where $\mu$ is a real positive number.
Also, in quantum field theory, the tachyonic field referred to often
happens in spontaneous symmetry breaking, where the field has an imaginary
mass due to the roll-in of the field towards a nonzero vacuum expectation
value caused by the effective mass term in the potential, which indicates
an instability. In this case, the evolution of the field is referred
to as tachyonic due to the resemblance to an imaginary mass. Moreover,
the idea of a combined subluminal-superluminal space could be explored
in quantum theories where quantum entanglement is coupled with space-time
manipulation, potentially leading to novel technologies in communication
or computation.

\end{document}